\documentclass[mathleft
]{an}
\usepackage{graphicx}
\usepackage{times}
\begin{document}

\Pagespan{789}{}
\Yearpublication{2006}%
\Yearsubmission{2005}%
\Month{11}%
\Volume{999}%
\Issue{88}%

\title{Towards a detection of individual g modes in the Sun}

\author{R.A. Garc\'\i a\inst{1}\fnmsep\thanks{Corresponding author:
  \email{rafael.garcia@cea.fr}\newline}
\and J. Ballot\inst{2}
\and A. Eff-Darwich\inst{3,4}
\and R. Garrido\inst{5}
\and A. Jim\'enez\inst{3,4}
\and S. Mathis\inst{1}
\and S. Mathur\inst{6}
\and A. Moya\inst{7}
\and P.L. Pall\'e\inst{3,4}
\and C. R\'egulo\inst{3,4}
\and D. Salabert\inst{3,4}
\and J.C. Su\'arez\inst{5}
\and S. Turck-Chi\`eze\inst{1}
}
\titlerunning{Towards detecting g modes in the Sun}
\authorrunning{R.A. Garc\'\i a et al.}
\institute{
Laboratoire AIM, CEA/DSM-CNRS, Universit\'e Paris 7 Diderot, IRFU/SAp-SEDI, Centre de Saclay, 91191, Gif-sur-Yvette, France
\and 
Laboratoire d'Astrophysique de Toulouse-Tarbes, Universit\'e de Toulouse, CNRS, F-31400, Toulouse, France
\and
Instituto de Astrof\'\i sica de Andaluc\'\i a (CSIC), Apartado 3004, 18080 Granada, Spain
\and
Universidad de La Laguna, 38206 La Laguna, Tenerife, Spain
\and
Instituto de Astrof'sica de Andaluc\'\i a CSIC, Camino. Bajo de Huetor, 50, Granada, Spain
\and 
High Altitude Observatory, Boulder, CO, 80302, USA
\and
Centro de Astrobiolog\'\i a (INTA-CSIC), Madrid, Spain
}

\received{}
\accepted{}
\publonline{later}


\abstract{Since the detection of the asymptotic properties of the dipole gravity modes in the Sun, the quest to find the individual gravity modes has continued. A deeper analysis of the GOLF/SoHO data unveils the presence of a pattern of peaks that could be interpreted as individual dipole gravity modes. The computed collapsed spectrum --around these candidate modes-- uncovers the presence of a quasi constant frequency splitting, in contrast with regions where no g modes are expected in which the collapsogram gives random results. Besides, the same technique applied to VIRGO/SoHO unveils some common signals between both power spectra. Thus, we can identify and characterize the modes, for example, with their central frequency and splittings. This would open the path towards new investigations to better constrain the solar core.}

\maketitle

\section{Introduction}
Our knowledge of the Sun's interior has been considerably improved by the observations of the solar acoustic oscillations (p modes). Indeed, accurate measurements of the low-degree p modes (Toutain et al. 1997; Thiery et al. 2000; Garc\'\i a et al. 2001; Salabert et al. 2004; Broomhall et al. 2009) have allowed us to infer the structure (e.g. Turck-Chi\`eze et al. 2001; Basu et al. 2009) and dynamics (Chaplin et al. 2001; Garc\'\i a et al. 2004; Mathis \& Zahn 2004, 2005; Turck-Chi\`eze et al. 2010) of the layers below the convective zone, from the radiative zone down to the solar core. However, the Sun's deepest regions are far from being fully known, and the inner radiative zone below 0.25~$R_{\odot}$, where about 50\% of the total mass is concentrated, will be only inferred by the detection of individual gravity (g) modes. For instance, few g modes would considerably improve the inversions of the rotational profile down to 0.1~$R_{\odot}$ (Mathur et al. 2008, 2009), while we also need to improve the detection of high-order low-degree p modes (Garc\'\i a et al. 2008c). The increasing background convective level towards the lower frequencies (e.g. Lefebvre et al. 2008), combined with very small amplitudes of those modes (several mm/s in the case of g modes, Belkacem et al. 2009) are the limiting factors for their detections. In the case of the low-degree, low-frequency p modes, accurate measurements are hardly obtained below 1~mHz (Bertello et al. 2000; Broomhall et al. 2008, 2010; Salabert et al. 2009), while no general consensus has been obtained for the detection of individual g modes (Appourchaux et al. 2010). It is necessary then to improve the current analysis methods and to develop new ones by trying to increase the signal-to-noise ratio (SNR) at low frequency. That can be achieved, for example, by reducing the convective noise (e.g. Garc\'\i a et al. 1999), by combining several instruments (e.g. Broomhall et al. 2007, 2010; Salabert \& Garc\'\i a 2008), or by using new instruments, such as PICARD (Thuillier et al. 2006), GOLF-NG (Turck- Chi\'eze et al. 2006) and the very promising HMI and AIA aboard SDO (Howe 2010).
Nevertheless, signals in the g-mode region have been detected in the GOLF and VIRGO data onboard SoHO, which are very unlikely to be due to instrument or solar noise (Jim\'enez \& Garc\'\i a 2009). Moreover, several patterns with high confidence level have been observed above the solar convective noise (Turck- Chi\`eze et al. 2004; Garc\'\i a et al. 2008a). The observation of the periodic signature of the g modes can be obtained by using methodologies applied in asteroseismology, where the large separations of solar-like stars with very small SNR can be measured even without being able to see clear features in the power spectrum (e.g. Mosser et al. 2009; Mathur et al. 2010). This technique has provided the first detection of the asymptotic periodicity of dipole g modes by computing the power spectrum of the power spectrum (PSPS) (Garc\'\i a et al. 2007). From this analysis, it was possible to infer an average speed in the solar core (3 to 5 times faster than the rest of the radiative zone), and an internal structure compatible with solar models based on the "old" abundances (Garc\'\i a et al. 2008b).

\section{Observations}

We have used 4472 days of Global Oscillations at Low Frequency (GOLF; Gabriel et al. 1995) time series calibrated into velocity (Pall\'e et al. 1999; Ulrich et al. 2000; Garc\'\i a et al. 2005), starting April 11, 1996. We have worked with a single -- full resolution -- power spectrum even knowing that GOLF has been observing in two different configurations with a different sensitivity to the visible solar disk (Garc\'\i a et al. 1998; Henney et al. 1999). To reduce the problem of the discretization in frequency, we have computed a 5 times zero-padded power spectrum (see the discussion concerning this problem in Gabriel et al. 2002). We have also used time series from the SPM/VIRGO package (Frohlich et al. 1995). 

\section{Data analysis}
In 2007, Garc\'\i a et al (2007) uncovered the existence of a pattern of peaks in the power spectrum of the GOLF instrument in the region between 25 and 140 $\mu$Hz that were equidistant in period. They interpreted this periodicity as the asymptotic properties of the dipole gravity modes. However, they were not able to distinguish the individual modes in the power spectrum. To face this problem and to try to identify the individual l=1 g modes we follow in this paper the same methodology commonly used in asteroseismology. Indeed, the first parameters that are searched for in asteroseismology are the frequency separations as p modes are equidistant in frequency (e.g. Bedding et al. 2010; Stello et al. 2010), while when the SNR is high enough, we can properly determine the characteristics of the individual p modes (e.g. Arentoft et al. 2008; Appourchaux et al. 2008; Garc\'\i a et al. 2009; Chaplin et al. 2010; Deheuvels et al. 2010). On the other hand, when the SNR of the modes is very small we are still able to determine the large separation even without seeing the individual p modes in the power spectrum (e.g. Mosser et al. 2009; Mathur et al. 2010). In such cases, the only thing that is generally done is to perform a heavy smooth of the power spectrum to unveil the p-mode hump in the region in which the frequency spacing is found. 

\begin{figure}[!htbp]
\includegraphics[width=5.3cm, trim = 1cm 1.2cm 2.5cm 3.5cm, angle=270]{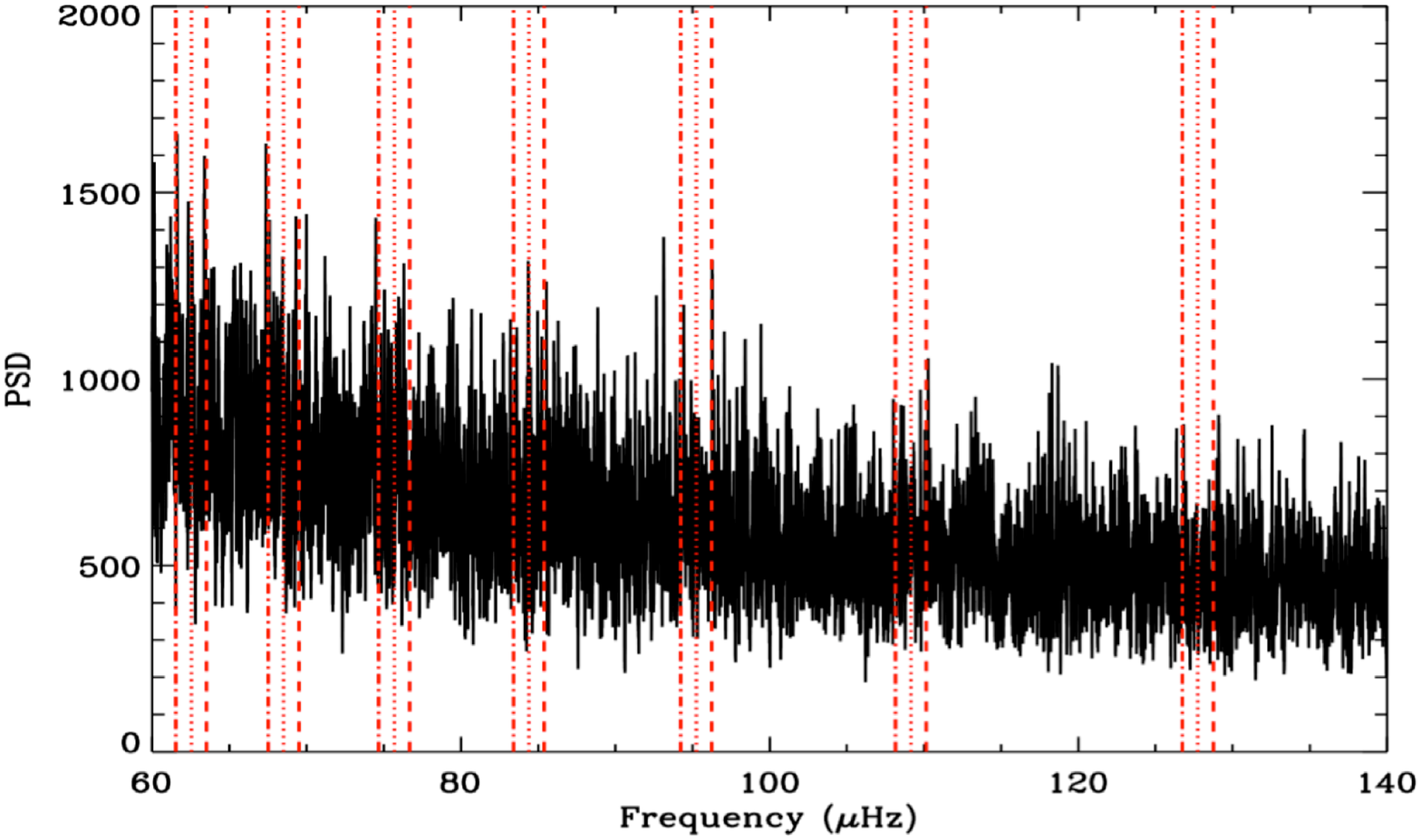}
\caption{5 times zero-padded PSD of 4472 days of GOLF velocity data in which a smooth by a boxcar of 41 nHz has been done. The vertical red dotted lines are the l=1 central (m=0) g-mode frequencies computed using the Saclay seismic model (Mathur et  al. 2007). Dashed and dot-dashed vertical red lines are the m=±1 components assuming a core rotating 4.5 times the rest of the radiative region.}
\label{fig1}
\end{figure}

\begin{figure}[!htbp]
\includegraphics[width=5.3cm, trim = 1cm 1.2cm 2.5cm 3.5cm, angle=270]{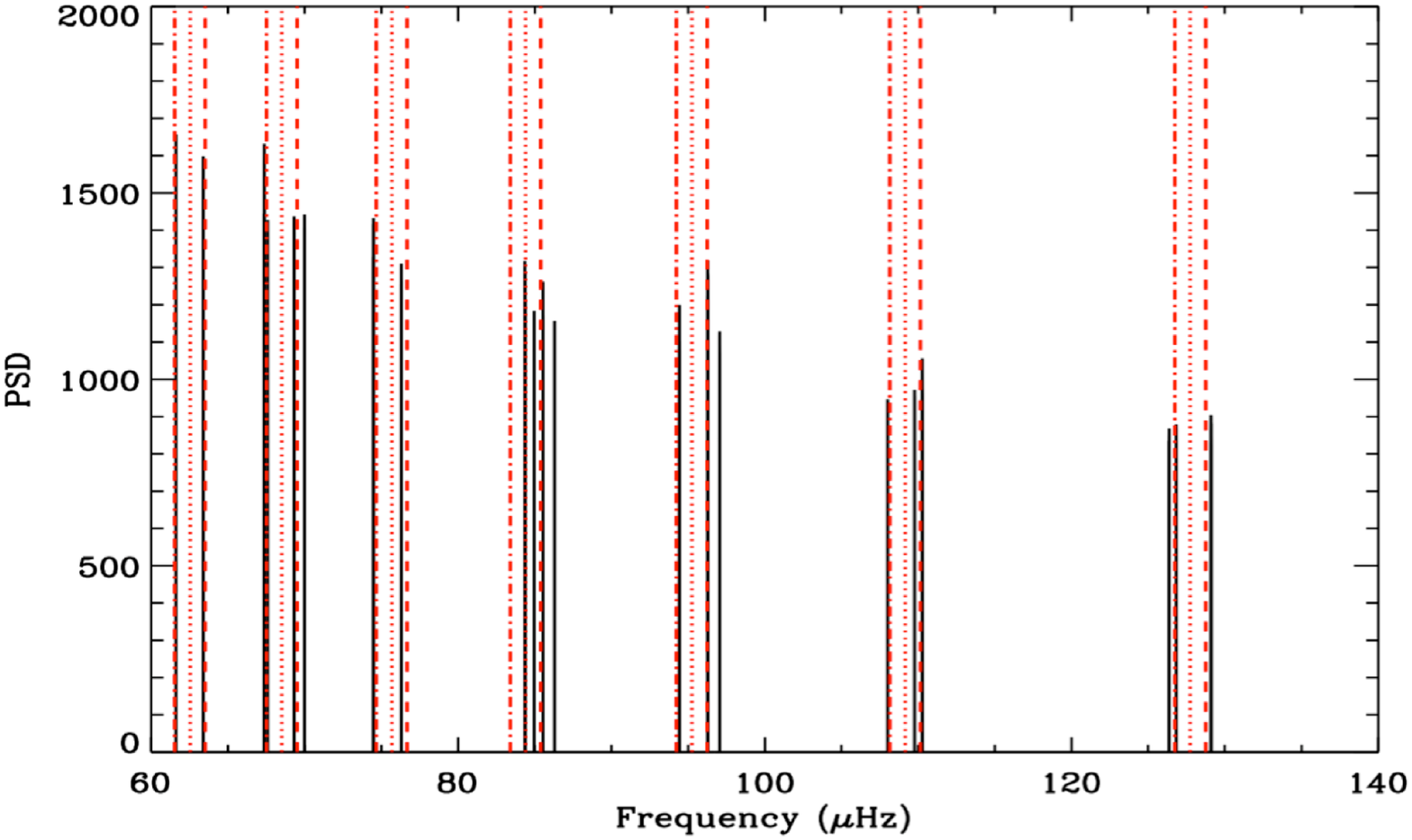}
\caption{Same than Fig.~1 but after filtering out all the peaks in the PSD in such a way that we only keep  the highest 0.3$\%$ of the  peaks in the PSD and also keeping the remaining peaks around $\pm$ 2.5 $\mu$Hz of the l=1 theoretical frequencies.}
\label{fig2}
\end{figure}

Figure~\ref{fig1} shows the 5 times zero-padded Power Spectral Density (PSD) of the GOLF time series. To increase the SNR of the possible signals we have applied a smooth of the spectrum by a boxcar function of 41 nHz. To guide the eyes, we have superimposed the predicted frequencies of the l=1, m=0 components of the dipole gravity modes (vertical red dotted lines) from the Saclay seismic model (Couvidat et al. 2003; Mathur et al. 2007). In this frequency range, the differences between this solar model, the model S (Christensen-Dalsgaard et al. 1996) and the M1 model from Nice (Provost et al. 2000) are less than 0.05 $\mu$Hz --which is inside the width of the vertical lines-- and we have also verified that the sensitivity of the pulsation codes to their parameters is negligible for the l=1 modes in this frequency range (Moya, Mathur, Garc\'\i a 2010). The dashed and dot-dashed vertical red lines are the m=$\pm$1 assuming a core rotating 4.5 times faster than the rest of the radiative region. This value is compatible with the range of the core rotation inferred by Garcia et al. (2007) and it matches the highest peaks of the PSD. Indeed, If we filter out all the peaks in such a way that we keep only the highest 0.3$\%$ of the peaks in the PSD and then we also remove the remaining spikes that are further than $\pm$2.5 $\mu$Hz of the predicted l=1 frequencies, we obtain the PSD showed in Fig.~\ref{fig2}. 

Now, we can clearly identify in the PSD the pattern of peaks --equidistant in period-- that was responsible of the detection of the $\Delta P_1$ spacing by Garc\'\i a et al. (2007) with more than 99.99$\%$ confidence level. Moreover, in 6 out of the 7 modes analyzed in this frequency region, the peaks seems to be split in frequency which is what we would have expected for gravity modes traveling in a region where the internal rotation is high. To perform a more quantitative analysis of the frequency splittings we compute the collapsograms.

\section{Collapsograms}
The so called m-collapsograms were originally developed for imaged instruments. This technique consists of shifting each m-component of the PSD by a quantity proportional to the splitting and co-adding all of them. The resultant average spectrum is fitted assuming a Lorentzian profile. The fitted peak with the smaller linewidth or maximum likelihood is the best result, which corresponds to a given displacement of the m-components: the searched splitting. The central frequency of the mode (m=0) is given by the frequency of the fitted Lorentzian (see Salabert et al. 2009 for a detailed explanation of the method).

\begin{figure}[!htbp]
\includegraphics[width=5.3cm, trim = 2cm 2.5cm 4.cm 3.5cm, angle=90]{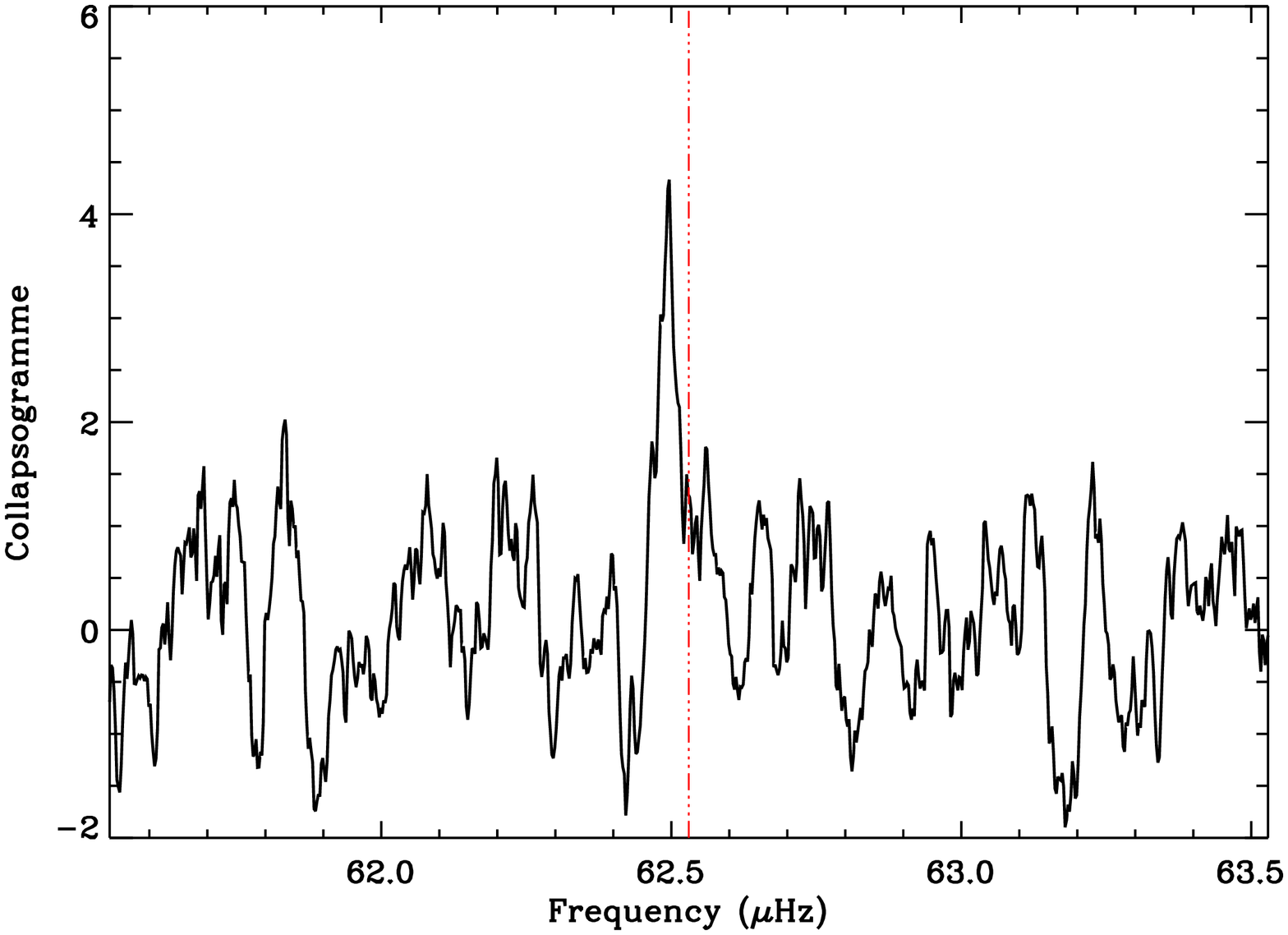}
\caption{Collapsed spectrum around the mode l=1, n=-10 showing the displacement giving the highest peak (895 nHz). The triple dot-dashed line marks the position of the predicted frequency.}
\label{fig3}
\end{figure}

In the case of GOLF, we only have one PSD in which only the even (l + m) components of the modes are visible. Thus, to mimic the methodology, we duplicate the GOLF PSD and we simply shift one of them in respect to the other one (like an auto correlation function). For each displacement we add both spectra and we look for the highest peak. The shift giving the highest peak in the collapsed spectrum provides the splitting and the frequency of the highest peak is the frequency of the mode. An example of the resultant collapsed spectrum is shown in Fig.~\ref{fig3} for the mode l=1, n=-10.

\begin{figure}[!htbp]
\includegraphics[width=8cm, trim = 0cm 4cm 1cm 5cm]{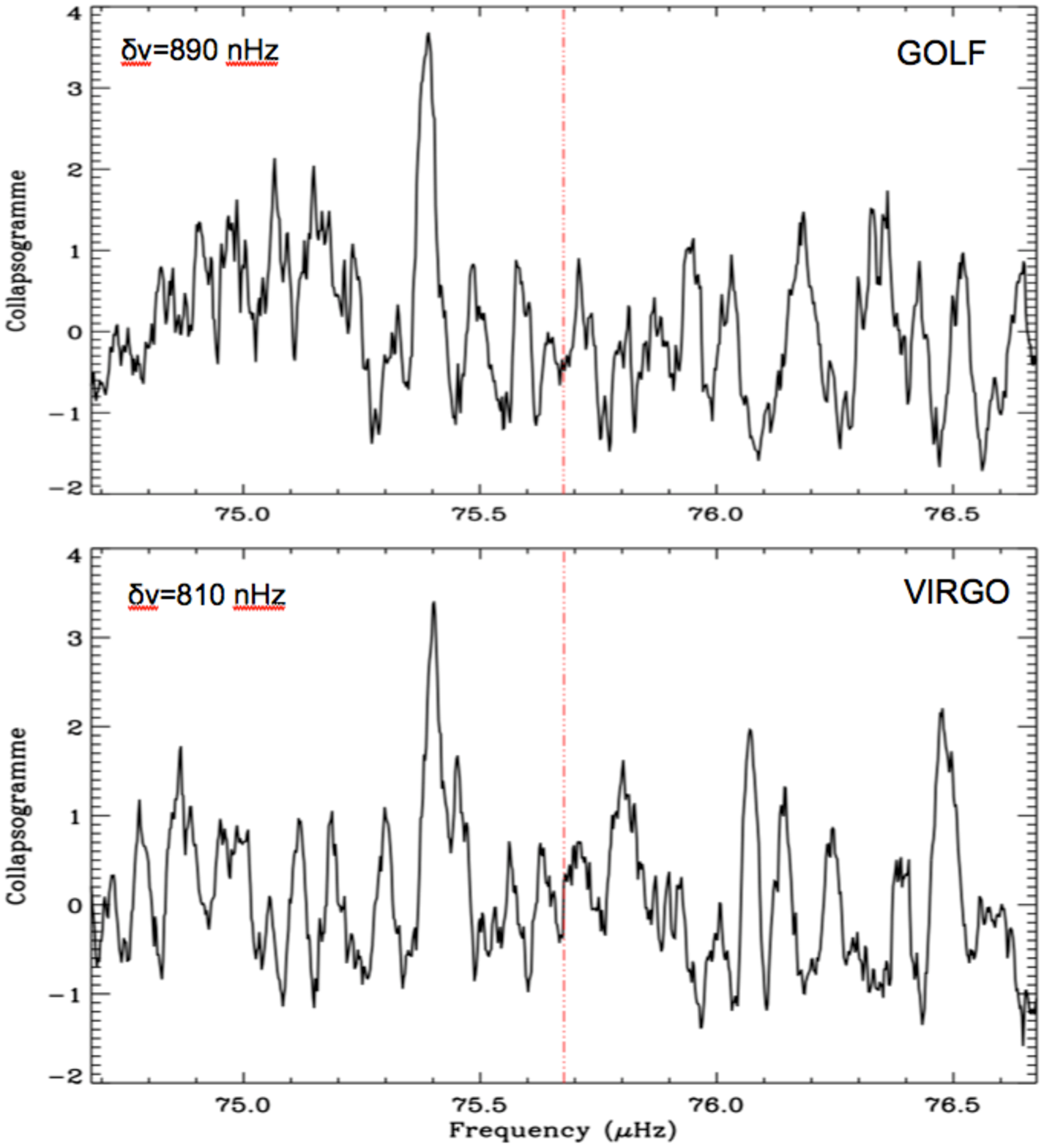}
\caption{Collapsed spectrum around the mode l=1, n=-8 for GOLF (Top) and VIRGO/SPM (Bottom).}
\label{fig4}
\end{figure}

Figure~\ref{fig4} shows the collapsed spectrum for the mode l=1, n=-8 for GOLF (top) and VIRGO (Bottom) data. In both cases the maximum is obtained at the same frequency but with slightly different frequency shift (splitting) of less than 10\% (2 frequency bins).

\begin{figure}[!htbp]
\includegraphics[width=5.9cm, angle=90]{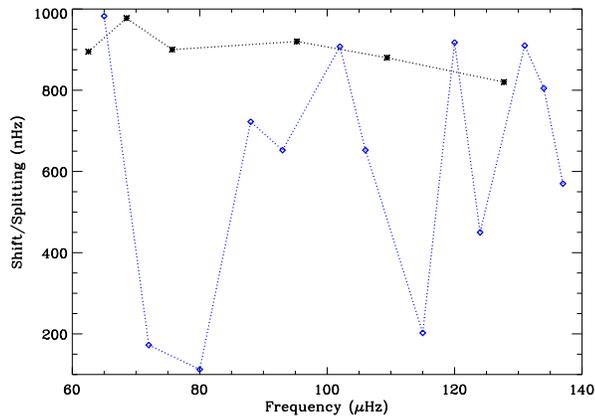}
\caption{Splittings obtained for the 6 out of 7 l=1 g modes in the regin 60-140 $\mu$Hz (Black asterisks). The blue diamonds ae the shifts obtained in regions where we do not expect gravity modes.}
\label{fig5}
\end{figure}

Figure~\ref{fig5} shows the resultant frequency shifts (splittings) of the collapsograms of 6 out of 7 dipole modes in the region 60 to 140 $\mu$Hz (black asterisks). A stable splitting around 850-900 nHz has been  found for all the modes. To do the collapsograms, we have scanned $\pm$1$\mu$Hz around the theoretical frequency, and rotation rates in a range 0.5 to 5 times the rotation of the rest of the radiative zone. Two modes were more difficult to characterize: the l= 1, n= -7  (~84$\mu$Hz) and the n= -5 (~109 $\mu$Hz) because the m=-1 and m=+1 components are not clearly visible simultaneously. For the second mode (n=-5), we managed to obtain a splitting in the same range than the other modes by taking the second highest peak in the collapsogram. The blue diamonds in Fig.~\ref{fig5} represents the collapsograms computed in regions in between the l=1 modes where only noise or high-degree g modes would be present.
A high dispersion of the results is found in the frequency shifts, which is what we expected for regions dominated by noise.

\section{Conclusions}
For the first time we have been able to identify and characterize the individual peaks responsible for the detection of the asymptotic $\Delta P_1$ spacing measured by Garc\'\i a et al. (2007) with more than 99.99$\%$ confidence level. Moreover, the peaks showed a quasi constant frequency splitting in a range 850 to 900 nHz.  This work is still in progress, but if these results are confirmed, it would allow us to better study the core of the Sun
\acknowledgements
GOLF and VIRGO instruments on board the SoHO spacecraft are a cooperative effort of many individuals, to whom we are indebted. SoHO is a space mission of international cooperation between ESA and NASA. D.S. acknowledges the support of the grant PNAyA 2007-62650 from the Spanish National Research Plan, A.M. the support from a Juan de la Cierva contract of the Spanish Ministry of Science and Innovation and J.B. the support through the ANR SIROCO. R.A.G. and S.T. thanks the support form CNES.

\end{document}